\documentclass[runningheads]{svmult}
\usepackage{makeidx}   % allows index generation
\usepackage{graphicx}  % standard LaTeX graphics tool
\usepackage{subeqnar}  % subnumbers individual equations
\usepackage{multicol}  % used for the two-column index
\usepackage{cropmark} % cropmarks for pages without
\usepackage{physprbb}  % centered layout of diverse elements, etc.
\begin{document}

\title*{The EDELWEISS Experiment: Status and Outlook}

\toctitle{The EDELWEISS Experiment: Status and Outlook}
\titlerunning{The EDELWEISS Experiment: Status and Outlook}

\author{J. Gascon\inst{1}
\and A. Benoit\inst{2}
\and A. Bonnevaux\inst{1}
\and L. Berg\'e\inst{3}
\and A. Broniatowski\inst{3}
\and B. Chambon\inst{1}
\and M. Chapellier\inst{4}
\and G. Chardin\inst{5}
\and P. Charvin\inst{5,6}
\and P. Cluzel\inst{1}
\and M. De J\'esus\inst{1}
\and P. Di Stefano\inst{5}
\and D. Drain\inst{1}
\and L. Dumoulin\inst{3}
\and G. Gerbier\inst{5}
\and C. Goldbach\inst{7}
\and M. Goyot\inst{1}
\and M. Gros\inst{5}
\and J.P. Hadjout\inst{1}
\and S. Herv\'e\inst{5}
\and A. Juillard\inst{5}
\and A. de Lesquen\inst{5}
\and M. Loidl\inst{5}
\and J. Mallet\inst{5}
\and S. Marnieros\inst{3}
\and O. Martineau\inst{1}
\and N. Mirabolfathi\inst{3}
\and L. Miramonti\inst{5}
\and L. Mosca\inst{5}
\and X.-F. Navick\inst{5}
\and G. Nollez\inst{7}
\and P. Pari\inst{4}
\and C. Pastor\inst{1}
\and E. Simon\inst{1}
\and M. Stern\inst{1}
\and L. Vagneron\inst{1}}

\authorrunning{Jules Gascon et al.}

\institute{IPNLyon-UCBL, IN2P3-CNRS, 4 rue Enrico Fermi, 69622 Villeurbanne Cedex, France
\and CRTBT, SPM-CNRS, BP 166, 38042 Grenoble, France
\and CSNSM, IN2P3-CNRS, Univ. Paris XI, bat 108, 91405 Orsay, France
\and CEA, Centre d'\'Etudes Nucl\'eaires de Saclay, DSM/DRECAM, 91191 Gif-sur-Yvette Cedex, France
\and CEA, Centre d'\'Etudes Nucl\'eaires de Saclay, DSM/DAPNIA, 91191 Gif-sur-Yvette Cedex, France
\and Laboratoire Souterrain de Modane, CEA-CNRS, 90 rue Polset, 73500 Modane, France
\and Institut d'Astrophysique de Paris, INSU-CNRS, 98 bis Bd Arago, 75014 Paris, France}

\maketitle

\begin{abstract}

The EDELWEISS Dark Matter search uses low-temperature
Ge detectors with heat and ionisation read-out
to identify nuclear recoils induced by elastic
collisions with WIMPs from the galactic halo.
Results from the operation of 70~g and 320~g Ge detectors
in the low-background environment of the Modane Underground Laboratory
(LSM) are presented.
\end{abstract}

\section{Introduction}

The EDELWEISS experiment is a direct WIMP
search, where nuclear recoils induced by
collisions with WIMPs from the galactic halo
are detected using Germanium detectors with
simultaneous measurement of ionisation and
phonon signals.
The comparison of the two signals makes possible
to separate on an event-by-event basis the
nuclear recoils from the electron recoils
induced by $\beta$- and $\gamma$- radioactivity
that constitute the major source of background
in most present-day direct WIMP searches.

The detectors are operated in the Laboratoire Souterrain de Modane
in the Fr\'ejus Tunnel under the French-Italian Alps.
The 1780 rock overburden (4800 m water equivalent) results in
a muon flux of about 4 m$^{-2}$day$^{-1}$ in the experimental
hall and the flux of neutrons in the 2-10 MeV range has been
measured to be
4$\pm$1 $\times$10$^{-6}$ cm$^{-2}$s$^{-1}$~\cite{bib-chazal}. 

The EDELWEISS-I phase consists in the operation of
one to three Ge detectors in the current one-litre cryostat,
their number being limited by the small volume.
In the year 2002, the program will enter a second phase
with the installation of a 100-litre cryostat
currently under construction,
allowing the use of up to 100 detectors.
In the mean time, the data-taking with the present-day
cryostat is devoted to the development of the detectors
and to setting improved limits on a possible WIMP
signal.

Outside the LSM, this activity is accompanied by
an intensive research and development program
aimed at improving the detector designs and
our understanding of their physical properties.
This includes work on phonon heat sensors using
NbSi thin films as Anderson insulators,
and the development of a facility to
calibrate detector responses to nuclear recoils
using a neutron beam and an array of NE213
scintillators to measure event-by-event the actual
nuclear recoil energy.

\section{The EDELWEISS detectors}

Important considerations in the design of
heat-and-ionisation detectors are size and
performance in terms of charge collection.
The imperfect charge collection of an electron recoil can
be mislead for the reduced ionisation yield of a nuclear recoil.
This must be avoided, e.g. by a careful electrode design
or by means of identification of events where the
charge has been deposited close to the detector surface.
A large detector size is clearly advantageous in terms of
event rate and surface-to-volume ratio.
However how this affects space-charge build-up and trapping
(affecting the ionisation signal) and how the increased
heat capacity affects the heat phonon signal requires
thorough investigations.
For these reasons, the EDELWEISS collaboration has tested
detectors with different sizes and electrode designs.

Two detector sizes have been tested: 70~g 
(48~mm diameter, 8~mm thick cylindrical Ge
monocrystals)~\cite{bib-distefano}
and 320~g (70~mm diameter, 20~mm thick)~\cite{bib-navick}.
The edges have been bevelled at an angle of 45$^o$.

The plane surfaces and wedges have been metallised for
ionisation measurement.
Two types of metallisation have been tested.
In the first one, the electrodes are boron and phosphorus implanted,
yielding a p-i-n structure.
In the second one, 100~nm Al layers are sputtered onto the surfaces
after etching.
In one of the 320~g prototypes the top electrode is
divided in a central part and a guard ring,
electrically decoupled for radial localisation of the
charge deposition.

The thermal sensor consists of a Neutron Transmutation Doped
germanium crystal (NTD) of a few mm$^3$ glued to the surface
of the detector.

\section{Results with 70~g detectors}

\begin{figure}[ht]
\begin{center}
\includegraphics[width=0.7\textwidth]{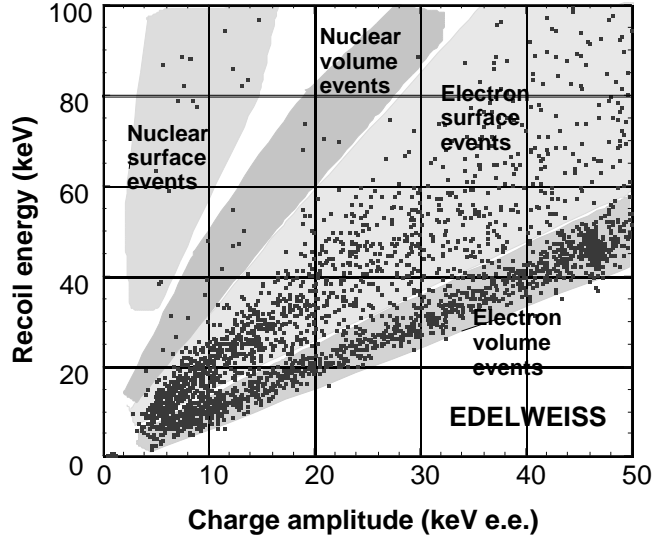}
\end{center}
\caption[]{Scatter diagram of the recoil energy vs. charge amplitude
for a 70~Ge detector (1.17 kg$\cdot$day exposure).
The four shaded areas correspond to the populations of events
attributed to (from left to right:) nuclear surface,
nuclear volume, electron surface and electron volume events.}
\label{fig-quatre} 
\end{figure}

The first results of the EDELWEISS collaboration
obtained in 1997 in its first test of a 70~g Ge detector
without the Roman lead shielding and radon removal
have been published recently~\cite{bib-distefano,bib-benoit}.
Resolutions of about 1 keV FWHM have been measured
on both ionisation and heat channels.
The ionisation trigger threshold was approximately 5 keV.
Fig.~\ref{fig-quatre} shows the scatter diagram of the recoil
energy versus the ionisation energy of events measured with
an exposure of 1.17 kg$\cdot$day~\cite{bib-benoit}.
As expected from calibrations with gamma and neutron sources,
electron recoil events appear along the diagonal and
the ionisation yield of nuclear recoil events is relatively
suppressed by a factor $\sim$3.5.
Electron recoil events with incomplete charge collection
fill the gap between the first two regions.
These events are attributed to interactions close to the surface of
the detector (mostly $\beta$ contamination and X rays) where
approximately half of the initially produced ionisation is lost.
There is no clear separation between the populations of nuclear
recoils and electron surface events.

A fourth population is observed: nuclear events with incomplete
charge collection, attributed to surface contamination from
alpha emitters.

The total event rate before electron recoil rejection
is 30 events/kg/day/keV in the 20 to 100 keV recoil energy range.
After rejection, the upper limit on the nuclear recoil event 
rate in that range is 0.6 event/kg/day/keV at 90\% C.L.

These results encouraged the pursuit of the project
with the construction of a neutron shield and 
improving the radiopurity of the detector environment.
A new implantation scheme for the p-i-n electrodes
was also tested on a new 70~g detector.
The cryostat was acoustically insulated from the rest of the
underground laboratory, 
with a copper mesh on the floor for better grounding.
An automatic system was set up to inverse at regular interval
for short periods the polarity on the electrodes of the Ge,
in order to free the trapped electrons that create undesirable
residual fields within the detector volume.
The protection against the radioactive background
has also been strenghtened.
A clean room was installed for handling the detectors. 
A 30~cm thick paraffin shielding against neutrons was installed.
Pure Nitrogen was circulated around the cryostat in order
to reduce radon accumulation.
All electronic components were moved away from the
detector and hidden behind the archeological lead shields.
A thorough selection of all materials entering
the experimental setup was instigated,
using the low-background counting facilities at the LSM.

\begin{figure}[ht]
\begin{center}
\includegraphics[width=0.6\textwidth]{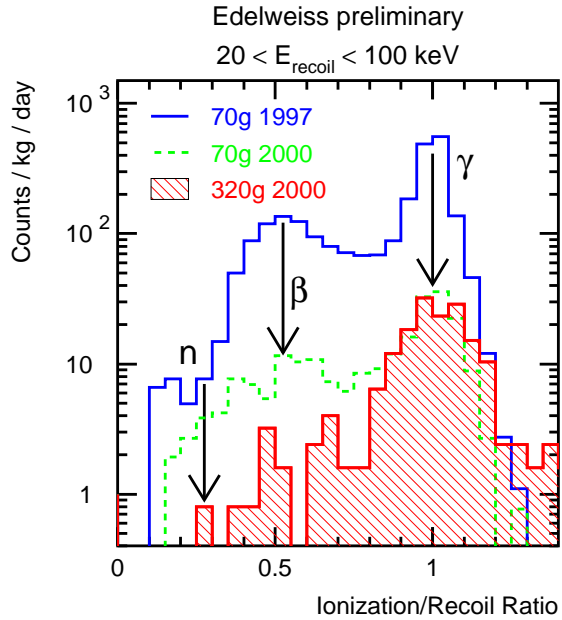}
\end{center}
\caption[]{Ratio of the ionisation yield to the recoil
energy for events with recoil energies between 20 and 100 keV.
The electron recoil yield has been normalised to 1 using
$\gamma$-ray calibration sources
and is expected to be approximately 0.3 and 0.5 for nuclear and
surface electron recoils, respectively.
The data set, normalised to 1 kg$\cdot$day, are:
(line) 1.8 kg$\cdot$day exposure for a 70~g detector in the 1997 configuration;
(dashed) 2.0 kg$\cdot$day exposure for a 70~g detector in the 2000 configuration
and (hatched histogram) 3.1 kg$\cdot$day exposure in the
center fiducial region of a 320~g detector with guard ring.}
\label{fig-qproj} 
\end{figure}

Tests with this new configuration were performed in
1999-2000.
A reduction of the overall background rate (before electron
recoils rejection) by a factor of ten was achieved,
as illustrated in fig.~\ref{fig-qproj}.
This figure shows the distribution of the ratio of
the ionisation yield to the recoil energy for recoil
energies between 20 and -100 keV range measured in 1997
(full line) and in a 1.97 kg$\cdot$day run with
the new configuration (dashed line).
This reduction is observed for electron recoils
with both complete (arrow labeled $\gamma$ on
fig.~\ref{fig-qproj}, with ionisation/recoil ratio $\sim$1)
and incomplete charge collection
(labeled $\beta$, ratio $\sim$0.5).

The efficiency-corrected rate of nuclear recoils
in the 20-100 keV range is 11$\pm$3 counts/kg/day.
It is only a factor two better than in the 1997
configuration, yielding an upper nuclear recoil rate of
0.25 nuclear counts/kg/day/keV in the 20-100 keV range
(90\% C.L.).
The subsequent test of a 320~g detector has
proven that this rate is essentially due to electron recoils
with improper charge collection.
The factor ten reduction of the rate observed for events
with a ionisation/recoil ratio close to 0.5 does not apply
for events with significantly worse charge collection (ratios $<$0.3).

\section{Results with a 320~g detector}

An important breakthrough came with the operation of
320~g Ge heat-and-ionisation detectors in the LSM,
the largest of this type of detectors in operation
in the world.
So far two detectors have been tested and up to
three should be installed at the end of the present run.

The most interesting results have been obtained with
a detector equipped with a guard ring electrode.
Work is still in progress in reducing the microphonic noise
on the ionisation and heat channels.
So far baseline resolutions of 2 keV on both channels have
been achieved.
The ionisation trigger threshold was kept under 7 keV over
an exposure time of 6.3 kg$\cdot$day, and consequently the
data analysis has been restricted to
nuclear recoils above 30 keV, a conservative estimate
of the effective threshold for these recoils.
The data taking is still under way.

\begin{figure}[htb]
\begin{center}
\includegraphics[width=0.6\textwidth]{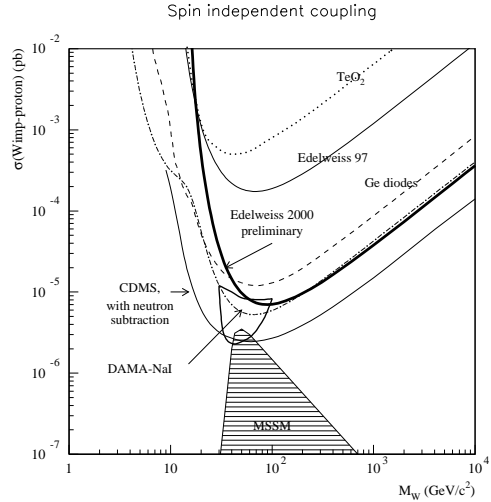}
\end{center}
\caption[]{WIMP exclusion limit (dark solid curve) obtained
from the preliminary data collected in the fiducial volume of a 320~g
EDELWEISS detector with guard ring electrode,
(3.1 kg$\cdot$day exposure).
Limits reported in Refs.~\cite{bib-dama}, \cite{bib-cdms}
and~\cite{bib-hdms} are also shown.} 
\label{fig-contour} 
\end{figure}

The fiducial region defined by rejecting events with
a significant signal on the guard electrode has been
estimated using a neutron calibration source and
it represents approximately 50\% of the total volume.
The distribution of the ratio of the ionisation yield
to the recoil energy obtained so far with an equivalent
exposure of 3.1 kg$\cdot$day is shown
in Fig.~\ref{fig-qproj} as a hatched histogram.
The overall rate before the rejection of electron recoils
with complete charge collection is comparable to the best
performance of the 70~g detectors.
More importantly, the rate of events with incomplete charge
collection is significantly reduced.
So far no nuclear recoils are observed in the 30 to 100 keV
recoil energy range, resulting in the preliminary exclusion
contour shown in fig.~\ref{fig-contour}.
It should be noted that this limit is obtained
without any neutron background subtraction,
and can be expected to improve as data taking is
progressing.

\section{Perspectives and Conclusions}

The present EDELWEISS 320~g detector is already setting interesting
WIMP limits.
As the limitation of the detectors have not been reached yet,
foreseeable improvements should come with the increase of statistics
in the current run.
Other improvements should arise with the better understanding of the
microphonics and other effects affecting the energy resolution and
threshold.
Before the end of 2001 the full 3$\times$320~g detector setup
should be installed.

In 2002 will start the installation of the 100-litre
EDELWEISS-II cryostat presently built in the CRTBT laboratory in
Grenoble. It will be able to accommodate up to 100
detectors and their electronics, providing the opportunity to
increase the sensitivity to a WIMP signal by more than two
orders of magnitude.

\section*{Acknowledgements}
The help of the technical staff of the Laboratoire Souterrain
de Modane and the participating laboratories is gratefully acknowledged.
This work has been partially funded by the EEC Network program under
contract ERBFMRXCT980167.

\end{document}